\documentstyle[twoside,fleqn,espcrc2,amsmath,cite,epsfig]{article}

\def\d{{\mathrm{d}}}

\def\cO#1{{\cal O}\left( {#1} \right)}

\def\as{\alpha_{\mathrm{s}}}

\def\cf{C_F}

\def\vk{{\vec k}}
\def\vn{{\vec n}}

\def\NP{{\mathrm{NP}}}
\def\PT{\mathrm{PT}}

\newcommand\cpc[3]{{{\it Comput. Phys. Commun. }{\bf #1} (#2) #3}}

\newcommand\jhep[3]{{{\it JHEP }{\bf #1} (#2) #3}}

\newcommand\npb[3]{{{\it Nucl. Phys. }{\bf B #1} (#2) #3}}

\newcommand\plb[3]{{{\it Phys. Lett. }{\bf B #1} (#2) #3}}

\newcommand\zpc[3]{{{\it Z. Physik }{\bf C #1} (#2) #3}}

\newcommand\epjdirc[3]{{{\it Eur. Phys. J. Direct }{\bf C #1} (#2) #3}}

\title{Power corrections and the interplay between perturbative and
  non-perturbative phenomena.\thanks{%
    Talk presented by GPS at QCD 99 Euroconference, Montpellier,
    July 1999.
    This work was supported in part by the EU
      Fourth Framework Programme `Training and Mobility of
      Researchers', Network `Quantum Chromodynamics and the Deep
      Structure of Elementary Particles', contract FMRX-CT98-0194 (DG
      12-MIHT).}  }

\author{
\vspace{-3.5cm}
\begin{flushright}
  Bicocca--FT--99--28\\
  hep-ph/9909324
\end{flushright}\vspace{2.0cm}
G.P. Salam\address{INFN, Sezione di Milano, 20133 Milano,
    Italy} and G. Zanderighi\address{Dipartimento di Fisica Nucleare e
    Teorica, Universit\`a di Pavia, and INFN Sezione di Pavia, 27100
    Pavia, Italy}}

\begin{document}
%----------------------------------------------------------------------

\begin{abstract}
  We discuss the issue of interplay between perturbative and
  non-perturbative phenomena for power corrections to $e^+e^-$ event
  shapes. 
\end{abstract}

\maketitle

%======================================================================
\section{Introduction}

For some time now it has been known that to describe event-shape
measures in $e^+e^-$ collisions, perturbative QCD on its own is not
sufficient. In general there is a need for significant additional
corrections, phenomenologically of the same size as the
next-to-leading corrections, of non-perturbative ($\NP$) origin.
Initial estimates for these corrections came from Monte Carlo event
generators \cite{Herwig,Jetset}. More recently, interest has arisen in
analytic approaches \cite{renormalons,DW,disp,twoloop}, based on
ideas such as the high-order behaviour of perturbation theory
(renormalons), or the concept of an infra-red-finite coupling, which
predict corrections of order $1/Q$, where $Q$ is the centre-of-mass
energy. This form is in rather good agreement with the experimentally
required corrections, but there is no a priori method (except perhaps
from the lattice) of determining the coefficient in front of $1/Q$: it
is a fundamentally non-perturbative quantity. There is however the
possibility that relative coefficients, from one observable to the
next, may be estimated. This is referred to as the \emph{universality}
hypothesis.

As will be discussed in section \ref{sec:trad}, traditional methods
for calculating the relative coefficients involve considering a $q\bar
q$ pair together with a single soft gluon. It turns out that event
shapes can be divided into two categories according to their behaviour
in the presence of a $q\bar q$ pair and many soft gluons: those whose
value is a linear combination of contributions from each of the soft
gluons, and those which instead involve a non-linear combination. The
traditional methods are adequate for linear observables. But they are
not suitable for non-linear ones because the presence of perturbative
($\PT$) soft gluons requires that one take into account correlations
between $\PT$ and $\NP$ gluons. Section~\ref{sec:pert} illustrates how
this is done.

%----------------------------------------------------------------------
\subsection{Event-shape definitions}
\label{sec:defs}

Let us first define the event-shapes that will be considered here. The
thrust is given by,
\begin{equation}
 T =  \max_{\vn} \frac{\sum_i |\vk_i.\vn|}{\sum_i |\vk_i|}\,,
\end{equation}
and measures the extent to which an event is `pencil-like'.  The
$C$-parameter is
\begin{equation}
\label{eq:Cdef}
  C = \frac32 \sum_{ij} k_i k_j \sin^2 \theta_{ij} \>.
\end{equation}
The heavy jet mass is,
\begin{equation}
   \rho_h = \frac{M_h^2}{Q^2}\,,
\end{equation}
where $M_h^2$ is the invariant mass of the heavier of the two
hemispheres defined by a plane perpendicular to the thrust axis $\vn$.
Finally we have the total jet-broadening,
\begin{equation}
\label{eq:Bdef}
  B_T=\frac{1}{2Q}\sum_i |k_{t,i}|\,,
\end{equation}
where the $k_t$ are measured perpendicular to the thrust axis. A
similar quantity, known as the wide-jet broadening, $B_W$, is defined
as the larger of the broadenings calculated separately in the two
hemispheres.

These events shapes have the property that in the limit of a two-jet
event, $1-T$, $C$, $\rho_h$, $B_T$ and $B_W$ go to zero.

%======================================================================
\section{Traditional power corrections}
\label{sec:trad}

There are many related approaches to the calculation of power
corrections to event shapes (and to a whole range of other
observables): renormalons \cite{renormalons}, the massive gluon
approach \cite{DW}, the dispersive approach \cite{disp}, the two-loop
approach \cite{twoloop}. In event-shape studies, they essentially
reduce to considering the behaviour of the observable in the presence
of the $q\bar q$ pair and a single soft gluon. For a soft gluon with
transverse momentum $k_t$ and rapidity $\eta$, the value of the event
shape variable $V$ is given by $k_t/Q$ times a `characteristic
function' $f_V(\eta)$:
\begin{center}
  \begin{tabular}{|r|c|c|c|c|c|} \hline
  $V=$  & $ 1-T$ & $C$ & $\rho_h$ & $ B_T$ & $B_W$ \\\hline
  $f_V=$ & $ e^{-|\eta|}$ & $\frac{3}{\cosh \eta}$ &  $ e^{-|\eta|}$ &
  $1$ & $1$ \\\hline 
  \end{tabular}
\end{center}
plus terms of order $k_t^2/Q^2$. Schematically, the power correction
can be seen as coming from the integral of this characteristic
function over rapidity and transverse momentum, including a factor of
$\as(k_t)$.  

More specifically, one considers only the non-perturbative part of
$\as$, (which we'll call $\alpha_\NP$), which is non-zero only for
small scales of the order of a GeV. So we have an expression for the
power correction which is\footnote{In practice one includes an
  additional factor of $2{\cal M}/\pi\simeq 1.14$, related to the
  splitting of the soft-gluon and details of the definition of
  $\as(k_t)$ \cite{twoloop,Milan}.}
\begin{equation}
\label{eq:defdV}
  \delta V = \frac{2\cf}{\pi}\int \frac{\d k_t}{k_t}\,
  \frac{k_t}{Q} \,\alpha_{\NP}(k_t) \int \d\eta\, 
  f_V(\eta) \,.
\end{equation}
Since all the observables have the same dependence on
$k_t$, the integral over $k_t$ yields the same value in each case:
\begin{align}
  \!\!\!\!\!\!  \!\!\!\!\!\!
  \int \frac{d k_t}{k_t} \frac{k_t}{Q} \alpha_{\NP}(k_t) &=
  \int^{\mu_I} \! d k_t\, \frac{\as(k_t)
   - \alpha_{\PT}(k_t)}{Q}  \nonumber \\
   &=\alpha_0(\mu_I) \frac{\mu_I}{Q} + \cO{\as(Q) \frac{\mu_I}{Q}},
\end{align}
where $\alpha_{\PT}$ is the perturbative expansion for $\as$ around
scale $Q$; the integral can be truncated at $\mu_I\sim 2$~GeV, because
beyond that scale $\alpha_{\NP}$ should die off very quickly.

The observable-dependent part of the power correction comes from the
integral over rapidity, which yields an observable-dependent
coefficient, $c_V$,
\begin{equation}
  \label{eq:cvdef}
  c_V = \int \d\eta\, 
  f_V(\eta) \,,
\end{equation}
multiplying $\alpha_0/Q$.

The postulate that $\alpha_0$ experimentally really is the same for
all variables, is referred to as universality. A priori we have no way
of calculating $\alpha_0$ (except perhaps on the lattice
\cite{lattice}), but we should be able to measure it in different
observables and find it to be consistently the same.

%======================================================================
\section{Interplay with the perturbative event}
\label{sec:pert}

The power correction as calculated above is valid for a situation in
which the non-perturbative gluon which we consider (termed `gluer' by
Dokshitzer \cite{gluer}) is the only particle present apart from the
initial $q\bar q$ pair. In practice however the event nearly always
contains many soft and collinear gluons of perturbative origin.  There
could also be several `gluers'.

So we need to understand what happens to the event shape in the
presence of more than one soft particle.

%----------------------------------------------------------------------
\subsection{The thrust and $C$-parameter}
\label{sec:TC}

Some variables, such as the thrust are simple. In the presence of many
soft particles, $1-T$ just becomes
\begin{equation}
  1-T \simeq \sum_i \frac{k_{t,i} }{Q}  e^{-|\eta_i|} \,,
\end{equation}
i.e.\ it is a linear combination of the contributions from each soft
particle. Thus the contribution of the gluer will just add to that of
the underlying perturbative event, and the consideration of the power
correction based on the presence of just a single gluon will remain
valid. A similar argument applies to the  $C$-parameter.

\begin{figure}[tb]
  \begin{center}
   \epsfig{file=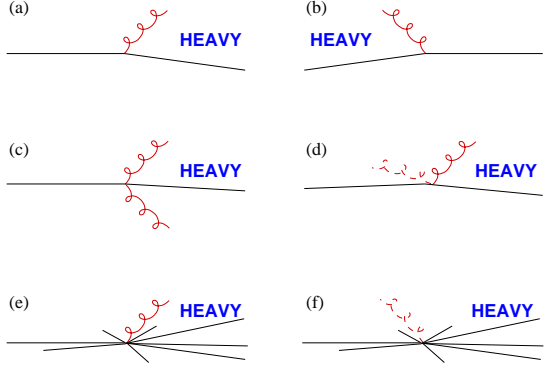,width=0.45\textwidth}
    \caption{Situations of relevance for the heavy-jet mass. Solid
      straight lines represent the underlying perturbative event.
      Curly lines are gluers. Dashed gluers do not contribute to the
      heavy-jet mass.}
    \label{fig:hjm}
  \end{center}
\end{figure}

%----------------------------------------------------------------------
\subsection{The heavy-jet mass}
\label{sec:hjm}

Considering the heavy-jet mass $\rho_h$, we find that the situation is
somewhat different. As given above, in the presence of a single gluon
(figs.\ \ref{fig:hjm}a and \ref{fig:hjm}b), $\rho_h$ is just $k_t/Q
e^{-|\eta|}$, because the heavy-jet is always the one containing the
gluon. In the presence of two particles in the same hemisphere (fig.\ 
\ref{fig:hjm}c), then the contribution is
\begin{equation}
  \frac{1}{Q}\left(k_{t,1}e^{-|\eta_1|} + k_{t,2}e^{-|\eta_2|} \right)\,,
\end{equation}
which is fine because it is a linear combination of the two
contributions. But when the two gluons are in opposite hemispheres
(fig.\ \ref{fig:hjm}d), the heavy jet mass is the larger of the two
hemisphere masses:
\begin{equation}
  \frac{1}{Q}\max\left(k_{t,1}e^{-|\eta_1|}, k_{t,2}e^{-|\eta_2|}
  \right)\,, 
\end{equation}
This is a non-linear combination of the two contributions, which means
that in the presence of the full $\PT$ event, the power correction
cannot simply be estimated by adding the 1-gluon result to the
perturbative answer.

So how do we determine the power correction in this case? The
fundamental point is to observe that the contribution to the event
shape from the underlying perturbative event is always much larger
than that from the gluers. To see why, consider that because of
Sudakov suppression of events without soft and collinear gluons, the
probability of an event-shape variable $V$ being smaller than some
value $V_{\max} \ll 1$ is roughly
\begin{equation}
  \label{eq:Vsud}
  P(V<V_{\max}) \sim \exp\left(-N \frac{\as\cf}{2\pi} \ln^2
    V_{\max}\right) 
\end{equation}
where $N$ is a variable-dependent number. Thus typical values for $V$
are of the order of 
\begin{equation*}
  V_{\mathrm{typical}} \sim \exp\left(-\frac{1}{\sqrt{\as}}\right)
  \sim \exp\left(-\sqrt{\ln Q}\right)\,. 
\end{equation*}
The contribution from non-perturbative effects,
\begin{equation*}
  V_{\NP} \sim \frac{1}{Q} \sim  \exp\left(-\ln Q\right)\,,
\end{equation*}
is much smaller.

As a result, in the limit of large $Q$, it will be the perturbative
event which `decides' which hemisphere is heavy (figs.\ \ref{fig:hjm}e
and \ref{fig:hjm}f). Once that information is known, $\rho_h$ is just
a linear combination of the contributions from all the particles in
the heavier hemisphere. So by invoking the large difference between
the typical perturbative and non-perturbative scales, we have been
able to reduce the heavy-jet mass to being linear in the contributions
from the gluers.  Since it is only gluers in one of the hemispheres
that are relevant, we halve the power correction compared to the
single gluon estimate \cite{AkhZak,Milan}.

%----------------------------------------------------------------------
\subsection{The jet broadenings}
\begin{figure}[tb]
  \begin{center}
    \epsfig{file=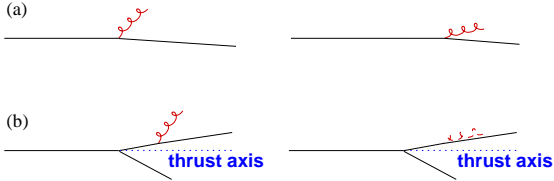,width=0.47\textwidth}
    \caption{Broadening for different quark axes.}
    \label{fig:B}
  \end{center}
\end{figure}

The jet-broadenings give a particularly rich example of a non-linear
event-shape observable which is reduced to being linear in the
presence of perturbative radiation.

In the presence of a single soft gluon there is a contribution from
the transverse momentum of the gluon and from the recoil momentum of
the quark. Noting the factor of a half in the definition of the
broadening, eq.\ \eqref{eq:Bdef}, one obtains that the contribution to
B is $k_t/Q$. In the presence of two soft gluons in the same
hemisphere, the situation becomes more complex: the contributions to
$B$ from the $k_t$'s of the soft gluons will simply be
$(k_{t,1}+k_{t,2})/2Q$; but the contribution from the quark recoil
will be $|\vk_{t,1}+\vk_{t,2}|/2Q$, which is anything but simple,
especially if one starts having to take into account many gluons. This
first complication is resolved by noting that recoil from perturbative
radiation will cause the quark to have a large (relative to the $\NP$
scale) transverse momentum $p_t$. After carrying out the integration
over the azimuthal angle of the gluer one obtains that the recoil due
to the gluer is
\begin{equation}
 \!\!\!\!\!\!\!\!\! \!\!\!\! 
 \int_0^{2\pi} \frac{d\phi}{2\pi} \left(\sqrt{p_t^2+k_t^2+2k_t
      p_t\cos\phi} - p_t\right) \simeq \frac{k_t^2}{4p_t}
\end{equation}
Being quadratic in $k_t$, this will lead at most to a $1/Q^2$
correction. So in the presence of $\PT$ radiation we can neglect quark
recoil. This halves one's expectation for the power correction (since
in the 1-gluon approximation equal contributions came the gluon and
from the quark recoil), but still leaves it proportional to $(\ln
Q/\Lambda)/Q$.  This form arises because the integral over rapidity,
\eqref{eq:cvdef} is bounded by the kinematic limit which is
approximately $\ln Q/k_t \simeq \ln Q/\Lambda$.  Experimentally, this
was found by both the H1 and JADE collaborations to be incompatible
with the data \cite{H1,JADE}. The JADE collaboration made the
additional observation that the power correction seemed to depend on
the value of $B$ itself.

The further element that has been neglected is that of the axis with
respect to which one measures the transverse momenta. One assumes a $d
k_t/k_t$ distribution for emission, where $k_t$ is given with respect
to the quark axis, but the broadening measures transverse momenta with
respect to the thrust axis. The extent to which these two axes
coincide affects the power correction. This is illustrated in
figure~\ref{fig:B}. If the quark axis and thrust axis coincide, fig.\ 
\ref{fig:B}a, then gluons with (the same $k_t$ and) large and small
angles to the quark axis contribute equally. If on the other hand the
quark axis does not coincide with the thrust axis, fig.\ 
\ref{fig:B}b, then gluons at large angles to the quark axis contribute
as before. But gluons close to the quark axis (angles smaller than the
angle between the thrust and quark axes) do not contribute --- or more
precisely their contribution is exactly cancelled by a corresponding
longitudinal recoil of the quark.  Therefore the rapidity integral for
the gluer does not extend beyond the quark rapidity, $\eta_q$, leading
to a power correction proportional to $\eta_q/Q$.  To determine
$\eta_q$, one observes that it is determined by the hardest gluon
(transverse momentum $p_t$) in the event, so that $\eta_q \simeq \ln
Q/p_t \simeq \ln 1/B$. Thus we arrive at the result that (for a single
hemisphere --- $B_1$), the coefficient of the power correction is
\cite{DMS}
\begin{figure}[tb]
  \begin{center}
    \epsfig{file=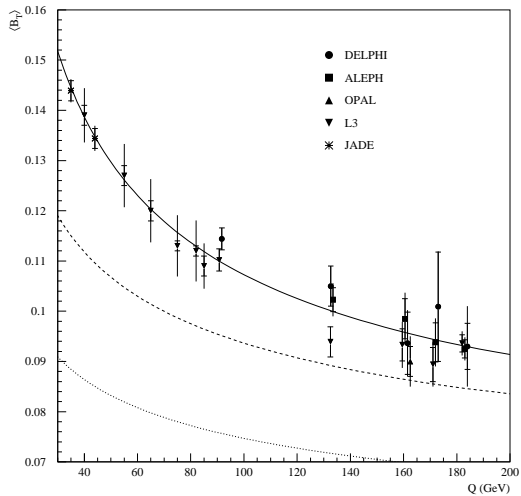,width=0.47\textwidth}
    \caption{Fit to the mean jet-broadening, using a power correction
      of the form \eqref{eq:cBtot}. The lowest line is the LO
      prediction, the next line includes the NLO corrections, and the
      highest line includes additionally the power correction.}
    \label{fig:meanB}
  \end{center}
\end{figure}
\begin{equation}
  \label{eq:BdepdB}
  c_{B_1}(B_1) = \frac12\ln \frac1{B_1} + \cO{1}\,.
\end{equation}
This applies separately for the broadenings from each hemisphere.

%......................................................................
\paragraph{Mean Broadenings}
Let us first use this information to determine the power correction to
the mean broadenings. We have to integrate the power correction with
the perturbative distribution for the broadening. This has to be done
separately for each hemisphere. The perturbative distribution for the
single-hemisphere broadening goes roughly as
\begin{equation}
  \label{eq:broaddist}
\!\!\!\!\!
\!\!\!\!\!
  \frac{B_1}{\sigma} \frac{\d \sigma}{\d B_1} \simeq 
  \frac{2\as \cf}{\pi} \ln \frac1{B_1} \exp\left(-\frac{\as \cf}{\pi}
    \ln^2 B_1\right) 
\end{equation}
Integrating this with \eqref{eq:BdepdB} and including a factor of two
to take into account the two hemispheres gives us the coefficient of
the power correction for the broadening:
\begin{equation}
  \label{eq:cBtot}
  \langle c_{B_T} \rangle = \frac{\pi}{\sqrt{\as\cf}} + \cO{1}\,,
\end{equation}
where the terms of $\cO{1}$ are to be found in \cite{DMS}. This form
for the power correction is used in figure~\ref{fig:meanB} where one
sees good agreement with the data.

For the wide-jet broadening, the distribution is similar to
\eqref{eq:broaddist} but with an extra factor of two in front of each
$\as$.  This together with the the fact that we have the $\NP$
contribution from just one hemisphere, gives a power correction
coefficient of
\begin{equation}
  \label{eq:cBwide}
  \langle c_{B_W} \rangle = \frac{\pi}{2\sqrt{2\as\cf}} + \cO{1}\,.
\end{equation}

\begin{figure*}[tb]
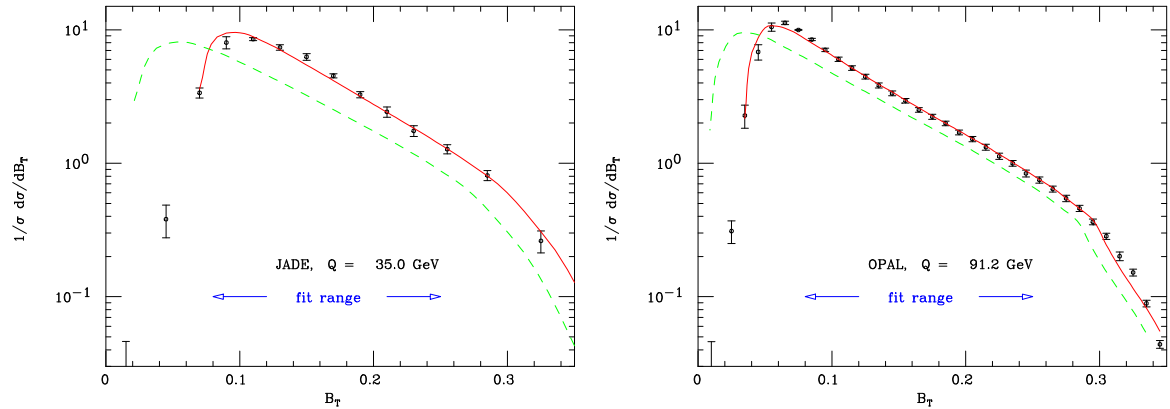

  \begin{center}
    \epsfig{file=btdist-jade35.eps,angle=90,width=0.47\textwidth}\quad
    \epsfig{file=btdist-opal91.eps,angle=90,width=0.47\textwidth}
    \caption{The distribution of the total
      jet-broadening from the JADE \cite{JADE} and OPAL \cite{OPAL}
      collaborations, compared to theoretical predictions: the dashed
      lines are the resummed perturbative prediction, while the solid
      lines include the power correction. The parameters, $\as=0.1158$
      and $\alpha_0=0.5368$, have been obtained from a fit to $B_T$
      distributions at a range of energies. }
    \label{fig:bdist}
  \end{center}
\end{figure*}

%......................................................................
\paragraph{Broadening distributions}
The situation for the distribution of the total jet-broadening is a
little trickier. Essentially, for small jet-broadenings each
hemisphere contains about half the total broadening and the power
correction is roughly twice \eqref{eq:BdepdB}:
\begin{equation}
  c_{B_T}(B_T) \simeq \ln \frac2{B_T} + \cO{1}\,.
\end{equation}
For larger jet-broadenings, most of the jet-broadening comes from one
hemisphere (whose power correction is just \eqref{eq:BdepdB}), while
the other hemisphere is essentially free to have almost any value of
the broadening and so its power correction is half of
\eqref{eq:cBtot} giving a total of
\begin{equation}
  c_{B_T}(B_T) \simeq \frac 12 \ln \frac1{B_T} +
  \frac{\pi}{2\sqrt{\as\cf}} + 
  \cO{1}\,. 
\end{equation}
Again, the full formulae, which interpolate between the two regimes,
are given in \cite{DMS}. Two examples of the broadening distribution
compared to data are shown in figure~\ref{fig:bdist} and the need for
and success of the power correction are clearly visible.

\begin{figure}[tb]
  \begin{center}
    \epsfig{file=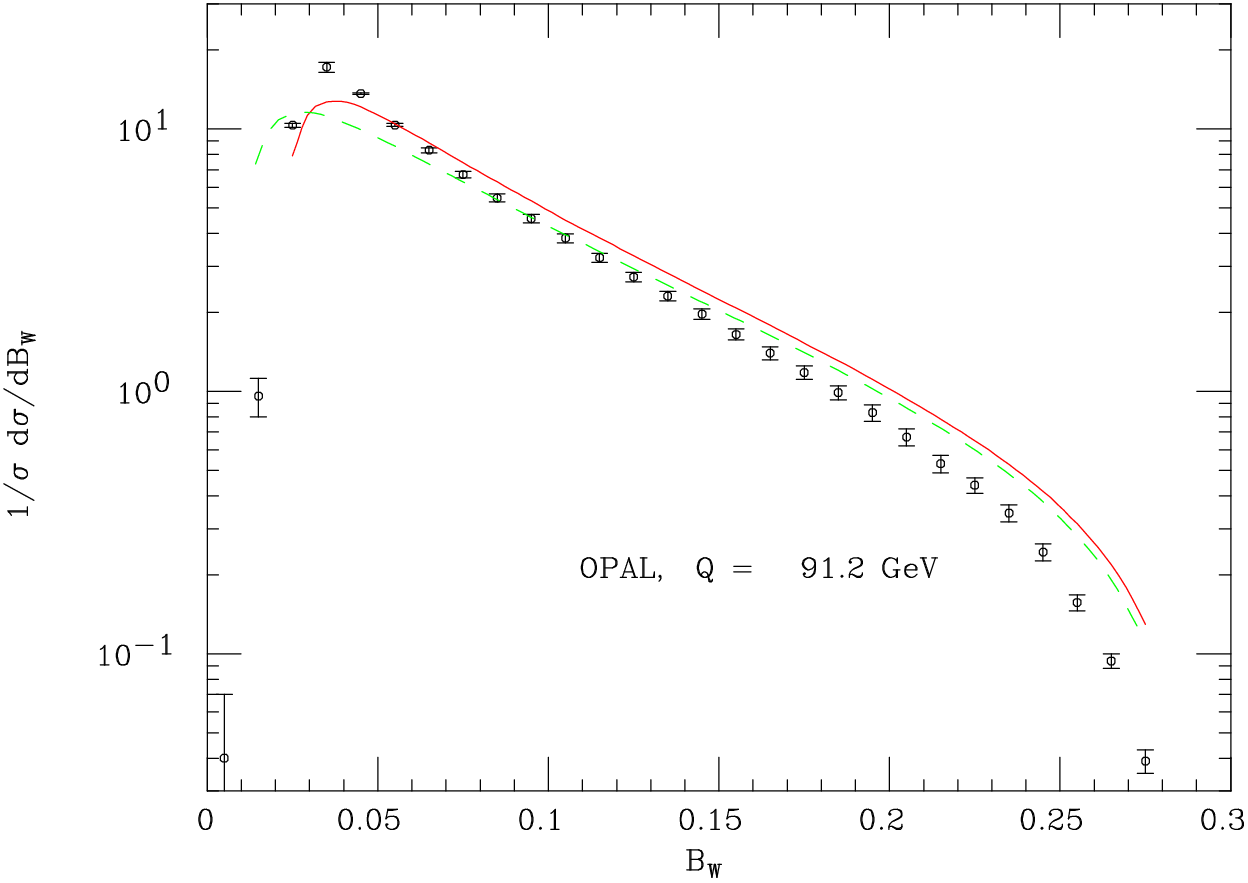,angle=90,width=0.47\textwidth}
    \caption{The distribution of the wide-jet broadening: a comparison 
      of OPAL data \cite{OPAL} and the resummed distribution, with
      (solid line) and without (dashed line) the power correction. The
      parameters used are those obtained from the fit to the $B_T$
      distribution, c.f.\ figure~\ref{fig:bdist}. (The best fit
      parameters for the $B_W$ distribution are quite different, and
      are shown in figure~\ref{fig:fits}).}
    \label{fig:bwdist}
  \end{center}
\end{figure}

The power correction to the distribution of the wide-jet broadening is
much simpler since we only have one hemisphere to worry about. It is
given simply by \eqref{eq:BdepdB}. The comparison to the data is
however quite unsatisfactory. The distribution is shown together with
OPAL data \cite{OPAL} in figure~\ref{fig:bwdist}. There is quite clear
disagreement, with a substantial need for the whole distribution to be
squeezed to smaller $B_W$, especially at moderate and large values of
$B_W$. Currently the origin of this problem is not understood. But it
should be noted that the wide-jet broadening is in some respects quite
a subtle variable: for example, going from a three-jet to a four-jet
event can reduce the value of $B_W$. These subtleties may translate
into large higher-order corrections, both to the perturbative and
non-perturbative parts.

\begin{figure}[tb]
  \begin{center}
    \epsfig{file=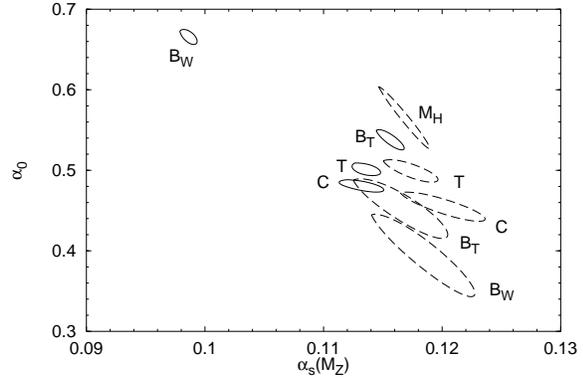,width=0.47\textwidth}
    \caption{2-$\sigma$ contours for fit results to various
      observables. No theoretical systematic errors are taken into
      account. Solid lines indicate fits to distributions, while
      dashed lines indicate fits to mean values. The solid curves for
      $C$ and $T$ are the results of fits carried out by the JADE
      collaboration \cite{JADE}. }
    \label{fig:fits}
  \end{center}
\end{figure}

%======================================================================
\section{Conclusions}
\label{sec:concl}

Figure~\ref{fig:fits} shows 2-$\sigma$ contours from fits to a variety
of observables, both for distributions and mean values. Given that one
does not necessarily expect $\alpha_0$ to be consistent between one
observable and another to better than $20\%$, because of potential
higher-order non-perturbative effects, and that theoretical systematic
errors have not been taken into account, the agreement for $\as$ and
$\alpha_0$ between observables is fairly reasonable. The only
exception is the wide-jet broadening, which as we have seen above has
yet to be fully understood.

Thus the approach of power corrections is a powerful tool in the
analysis of event shapes. An important element in its success is a
treatment of the interplay between perturbative and non-perturbative
radiation.

%======================================================================
\section*{Acknowledgements}
We are grateful to Mrinal Dasgupta, Yuri Dokshitzer and Pino
Marchesini for helpful discussions. One of us (GPS) would also like to
thank Pedro Movilla Fern\'andez for pointing out a bug in one of the
fitting programs.

%======================================================================


\begin{thebibliography}{99}

\bibitem{Herwig} G. Marchesini, B.R. Webber, G. Abbiendi, I.G.
  Knowles, M.H. Seymour and L. Stanco, \cpc{67}{1992}{465}.

\bibitem{Jetset} T. Sjostrand, \cpc{82}{1994}{74}.

\bibitem{renormalons} 
A.V. Manohar and M.B. Wise, \plb{344}{1995}{407} [hep-ph/9406392];
G.P.~Korchemsky and G.~Sterman,
%``Nonperturbative corrections in resummed cross-sections,''
\npb{437}{1995}{415} [hep-ph/9411211];
R.~Akhoury and V.I.~Zakharov, \plb{357}{1995}{646} [hep-ph/9504248].
%``Leading power corrections in QCD: From renormalons to
 %phenomenology,''

\bibitem{DW}
 Yu.L.\ Dokshitzer and  B.R.\ Webber,  \plb{352}{1995}{451}.

\bibitem{disp} M.\ Beneke and V.M.\ Braun, \plb{348}{1995}{513}
  [hep-ph/9411229] ;
 P.\ Ball, M.\ Beneke and V.M.\ Braun,  \npb{452}{1995}{563}
 [hep-ph/9502300] ;
 Yu.L.\ Dokshitzer, G. Marchesini and  B.R.\ Webber,
 \npb{469}{1996}{93} [9512336].

\bibitem{twoloop}   Yu.L. \ Dokshitzer, A.\ Lucenti,  G.\ Marchesini
  and G.P.\ Salam,  \npb{511}{1998}{396} [hep-ph/9707532].

\bibitem{Milan}  Yu.L.\ Dokshitzer, A.\ Lucenti, G.\ Marchesini and
  G.P.\ Salam, \jhep{05}{1998}{003} [hep-ph/9802381].


\bibitem{lattice} G. Burgio, F. Di Renzo, C. Parrinello and C.
  Pittori, hep-ph/9808258.
  
\bibitem{gluer}
  Yu.L.\ Dokshitzer, V.A.\  Khoze, A.H.\ Mueller and S.I.\ Troyan, {\em
    Basics of Perturbative QCD}, ed. J. Tran Thanh Van, Editions
  Fronti\`eres, Gif-sur-Yvette, 1991.

\bibitem{AkhZak} R. Akhoury and V.I. Zakharov, \npb{465}{1996}{295}
  [hep-ph/9507253].


 
\bibitem{H1} H1 Collaboration, C.\ Adloff et al., \plb{406}{1997}{256}
  [hep-ex/9706002]; contribution 530 to ICHEP July 1998, Vancouver,
  Canada.

\bibitem{JADE} P.A.\ Movilla Fern\'andez, talk at QCD Euroconference,
  Montpellier, France, July 1998,
  hep-ex/9808005;\\
  P.A.\ Movilla Fern\'andez, O.\ Biebel and S.\ Bethke, paper
  contributed to ICHEP-98, Vancouver, Canada, hep-ex/9807007.
  
\bibitem{DMS} Yu.L.\ Dokshitzer, G.\ Marchesini and G.P.\ Salam,
  \epjdirc{3}{1999}{1} [hep-ph/9812487].
  

\bibitem{OPAL}  OPAL Collaboration, P.D.\ Acton et al.,
  \zpc{59}{1993}{1};
  




\end{thebibliography}
\end{document}